\documentclass[%
reprint,
superscriptaddress,
amsmath,amssymb,amsthm,
aps,
soul
]{revtex4-2}

\usepackage{hyperref}
\hypersetup{
  colorlinks   = true, %Colours links instead of ugly boxes
  urlcolor     = blue, %Colour for external hyperlinks
  linkcolor    = blue, %Colour of internal links
  citecolor   = blue %Colour of citations
}

\usepackage{graphicx}% Include figure files
\usepackage{dcolumn}% Align table columns on decimal point
\usepackage{bm,soul}% bold math
\usepackage{xcolor}

\usepackage{color}

\begin{document}

\preprint{APS/123-QED}

\title{Implementation of Trained Factorization Machine Recommendation System on Quantum Annealer}% Force line breaks with \\

\author{Chen-Yu Liu}
\email{cyliuphys@gapp.nthu.edu.tw}
\affiliation{Hon Hai (Foxconn) Research Institute, Taipei, Taiwan}
\affiliation{Graduate Institute of Applied Physics, National Taiwan University, Taipei, Taiwan}

\author{Hsin-Yu Wang}
\email{b07703009@ntu.edu.tw}
\affiliation{Department of Finance, National Taiwan University, Taipei, Taiwan}
\affiliation{Department of Mathematics, National Taiwan University, Taipei, Taiwan}

\author{Pei-Yen Liao}
\email{b08705006@ntu.edu.tw}
\affiliation{Department of Mathematics, National Taiwan University, Taipei, Taiwan}

\author{Ching-Jui Lai}
\email{cjlai72@mail.ncku.edu.tw}
\affiliation{Department of Mathematics, National Cheng Kung University, Tainan, Taiwan}

\author{Min-Hsiu Hsieh}
\email{min-hsiu.hsieh@foxconn.com}
\affiliation{Hon Hai (Foxconn) Research Institute, Taipei, Taiwan}

\date{\today}% It is always \today, today,
             %  but any date may be explicitly specified

\begin{abstract}
 
Factorization Machine (FM) is the most commonly used model to build a recommendation system since it can incorporate side information to improve performance. 
However, producing item suggestions for a given user with a trained FM is time-consuming. To address this problem, we propose a quadratic unconstrained binary optimization (QUBO) scheme to combine with FM and apply quantum annealing (QA) computation. Compared to classical methods, this hybrid algorithm provides a fast sub-optimal sampling of good user suggestions. We then demonstrate the aforementioned computational behavior on current noisy intermediate-scale quantum (NISQ) hardware by experimenting with a real example on a D-Wave annealer. 

\end{abstract}

%\keywords{Suggested keywords}%Use showkeys class option if keyword
                              %display desired
\maketitle

%\tableofcontents

\section{INTRODUCTION}

%** A general introduction to the Recommendation system **

Recommendation systems are widely used to predict the rating or preference of items for arbitrary users, such as suggesting books or  movies, and also for many other business applications \cite{rs1, rs2, rs3, rs4}. A  common principle for constructing a recommendation system assumes that users prefer similar things based on past behavior. Amongst several approaches, the objective of Matrix Factorization (MF) is to find the factor matrices of historical data \cite{cf1, mf1}. Together with MF, other improvements such as SVD++ \cite{mf2}, pairwise interaction tensor factorization (PITF) \cite{mf3}, factorizing personalized Markov chains (FPMC) \cite{mf4} and Monte Carlo on bipartite graphs  \cite{rw1} has also been comprehensively studied. These methods introduce specialized models to treat specific tasks but with the trade-off of losing generality.

% Why we choose FM 
To fix this, the Factorization Machine (FM) \cite{fm1} is proposed as a more general supervised learning method, unifying MF, SVD++, PITF, and FPMC. With the capability to input arbitrary real-valued feature vectors to the model, including side information (e.g., gender and age), FM can be used for regression, classification, and ranking tasks. Moreover, it can be trained with linear complexity and accurately estimate model parameters from a sparse dataset, which makes FM highly competent for analyzing large datasets. Thus, it is more natural and adequate to utilize FM for real-world applications than the other methods mentioned earlier.

% Introduction to QML and its drawback
In recent years, quantum computing has become one of the most popular domains, which aims to benefit from the superposition nature of quantum states. Quantum Machine Learning (QML) employs parameterized quantum circuits as a neural network, which can then be applied to different classes of machine learning, such as Quantum Neural Network (QNN) \cite{qnn1, qnn2, qnn3}, Quantum Convolutional Neural Network (QCNN) \cite{qcnn1, qcnn2, qcnn3}, Quantum Reinforcement Learning (QRL) \cite{qrl1, qrl2, qrl3}, and Quantum Generative Adversarial Neural Network (QGAN) \cite{qgan1, qgan2, qgan3, qgan4, qgan5}. Besides the structural exploration, several works \cite{qnnlearn1, qnnlearn2, qnnlearn3, qnnlearn4, qnntrain1, qnntrain2, qnntrain3, qnntrain4, qnntrain5} have also investigated the learnability and trainability of QNN. Although the data encoding to a Hilbert space may give potential computational advantages, at the current stage of the NISQ era, only low-dimensional or small sample problems can be implemented on real-world hardware. Thus it is hard to justify the advantage experimentally, and this needs to be further studied \cite{qnn4}. 

% Introduction to QRS and QIRS
The concept of a Quantum Recommendation System (QRS) has also been proposed \cite{qrs1}, for which the idea is to sample an approximation of the preference matrix by quantum singular value estimation and quantum projection. In that framework, QRS takes $O(\text{poly}(k)\text{polylog}(mn))$ running time with $k$ the rank of the approximation and $mn$ the matrix dimension, while classically reconstructing an approximation of the preference matrix requires polynomial time $mn$. 
However, a classical analog of QRS, called Quantum-Inspired Recommendation System (QIRS), provides $O(\text{poly}(k)\text{log}(mn))$ running time \cite{qirs2}, which closes the gap between the classical and quantum methods. Besides gate-based quantum computing, it is possible to apply quantum annealing for feature selection \cite{fsqc1} and use these features to train a classical ItemKNN content-based model \cite{itemknn1}. This hybrid QPU solver shows good scalability for larger instances but may not have a significant advantage in the running time. 

% limitation QRS and QIRS: side information 
The performance of QRS or QIRS depends on a good low-rank approximation to the preference matrix \cite{qrs1, qirs2}. Moreover, since these algorithms do not incorporate side information, they may not be competitive with FM. Thus in the recommendation system domain, an approach to both utilize the side information and attain potential speedup is a crucial target for practical quantum computing applications. 

% This work 

In this work, we formulate a hybrid recommendation system by incorporating a classically trained FM with a quadratic unconstrained binary optimization (QUBO) scheme to be solved quantumly. Applying Quantum Annealing (QA) computation enables us to sample a large amount of sub-optimal suggestions in a short period of time compared to classical methods.  Our algorithm will soon provide a computational advantage from the prospects of today's developing NISQ hardware, such as a D-Wave annealer, for practical problems.

\subsection{Main Results} 
% At least half page
Our main results in a both theoretical and experimental speedup of a QA-assisted FM recommendation system are summarized as follows:
\begin{itemize}
    \item Suggestion process as energy minimizing task: We propose a QUBO scheme for the suggestion process of the FM recommendation system. The energy minimization task of the corresponding Ising problem is equivalent to finding the highest score (rating) candidates in the recommendation system.
    
    \item NISQ advantage of Quantum Annealing in recommendation system: 
    Ideally, the quantum annealing process aims to yield the ground state of the Hamiltonian. However, the presence of noise and the limitation of annealing time on current NISQ hardware often leads to a multitude of sub-optimal solutions for the Hamiltonian. Interestingly, these sub-optimal solutions align quite well with the objectives of the recommendation system, which is designed to provide multiple suggestions rather than just the absolute best one.
 
    \item Scalability: The capability of solving fully-connected size-$N_p$ Ising problems of the most advanced D-Wave Advantage 4.1 system is $N_p = 145$. At the same time, the corresponding QUBO size of suggesting $N_m$ candidates is $\lceil \log_2 N_m \rceil$ in our formulation. Hence, theoretically, the solvable problem size $N_m$ scales up to $2^{145} \approx 10^{43}$ in today's quantum hardware.
    
\end{itemize}

\subsection{Related Work}
% structure fixing 
% 1. recommendation system 
% 2. Factorization machine 
% 3. QUBO 
% Related works and what is the difference between our work and their work

Several works apply the QUBO scheme to the machine learning study. Date \textit{et al.} \cite{fmqubo4} proposes QUBO formulations of linear regression, support vector machine (SVM), and balanced $k$-means clustering, where the required qubit number usually scales as $O(N_{\text{data}}^2)$, with $N_{\text{data}}$ the number of the data points of the training data. However, $N_{\text{data}}$ could go easily beyond $10^3$ in most machine learning applications, which is higher than the capability of the most advanced quantum system for solving an Ising problem: $N_{\text{data}}=145$ from D-Wave \cite{dwave_upper_limit}. One can transform an FM into a QUBO problem by binary encoding with a size equivalent to the length of the input vector. The works \cite{fmqubo1, fmqubo2, fmqubo3, fmqubo5} have applied this idea in structural design problems to maximize the acquisition function associated to an FM. In this manner, the size of the QUBO problem is usually within a range that today's hardware could reach. These examples open the possibility of applying such a QUBO-FM scheme to any FM-related research.

\section{Preliminaries}
In this preliminary section, we review the key ingredients of our work: FM, QUBO, and Quantum Annealing (QA). FM is the most commonly used model in recommendation systems and machine learning. The QUBO scheme can apply to various combinatorial optimization problems, and QA is an effective optimization method for solving QUBO problems.
%%%%%%%%%%%%%%%%%%%%%%%%%%%%%%%%%%%%%%%%%%%%%%%%%%%%%%%
\subsection{Factorization Machine}
\label{sec:fm}

A factorization machine (FM) is a supervised learning algorithm that can be used for classification, regression, and ranking tasks \cite{fm1}. Here we consider FM with degree $D = 2$, which involves only single and pairwise interactions of items. The model equation for an FM of degree $D = 2$ with variables 
$$\vec{x}=(x_1,\dots,x_d)$$
 is defined as:
\begin{eqnarray}
\label{eq:fm}
y_{\text{fm}}(\vec{x}) := w_0 + \vec{w}\cdot\vec{x} + \sum_{i<j}^d v_i^T v_j x_i x_j,
\end{eqnarray}
where $w_0 \in \mathbb{R}$ is the global bias, $\vec{w}=(w_1,\dots,w_d)\in \mathbb{R}^d$ refers to the weights of each variables and 
\begin{eqnarray}
{\bf V}=\left[
  \begin{array}{cccc}
    \vrule & \vrule & & \vrule\\
    v_{1} & v_{2} & \ldots & v_{d} \\
    \vrule & \vrule & & \vrule 
  \end{array}
\right]
 \in \mathbb{R}^{k \times d}
\end{eqnarray} 
denotes the feature embeddings with $k$ the dimensionality of latent factors. The term $v_i^T v_j$ represents the interaction between the $i$-th and $j$-th terms.

We can use FM to construct a recommendation system: Separate the variables $\vec{x}$ into  $x_i = u_i$ for $1 \le i \le n_u$ and $x_{i+n_u} = m_i$ for $1 \le i \le n_m$, so that the vector $\vec{u}=(u_1,\dots, u_{n_u}) $ denotes the information of a user and $\vec{m}=(m_1,\dots, m_{n_m})$ represents the information of an item to be recommended, with the dimension relation $n_u + n_m = d$, this then turns $y_{\text{fm}}(\vec{x})$ into a predictor that estimates the score or rating for such a $(\vec{u}, \vec{m})$ pair. 

FM model is known in the original FM paper \cite{fm1} to have linear complexity, the capability of parameter estimation under sparse data, and the ability to work with any real-valued feature vector. As a result, one can use an FM model as a more general predictor than other state-of-art factorization models that only work on restricted input data and require individual task analysis. Furthermore, it has been shown in the same article that FM can mimic biased MF, SVD++, PITF, and FPMC, which indicates that FM is user-friendly for non-experts wanting to work with factorization models. 

One can describe a typical use case of FM as follows:
\begin{enumerate}
    \item Prepare a set of data with user information, item information, and ratings (or scores) of the items produced by the user.
    \item With such a dataset, $w_0$, $w$, and $v$ can then be learned by optimization algorithms, such as stochastic gradient descent (SGD), alternating least-squares (ALS) optimization, as well as Bayesian inference using Markov Chain Monte Carlo (MCMC) \cite{fm2}. This step is also called the training process.
    \item After training an FM, the predictor $y_{\text{fm}}(\vec{x})$ can estimate the ratings of every item in the dataset with given user information.
\end{enumerate}

% Description of suggestion process detail, what is the definition ? 
% elaborate more 
% k in independent of the system size, should be ignored in big O notation, here we focus on the effect of system size. 
With a trained FM, we perform the suggestion process of a given user in the following manner. First, with $N_m$ items in the dataset, each item is input to the FM to produce the corresponding rating. Next, we sort the obtained $N_m$ ratings and provide the list of the top $k_s$ suggestions to the user, where $k_s$ is the pre-determined number of suggestions.

In the rest of this work, the suggestion method that calculates all ratings directly and then sorts is called the ``direct method''. 

For a trained FM with $k$ the dimensionality of the factorization, the complexity of finding the predictor $y_{\text{fm}}(\vec{x})|_{\vec{u}=\vec{u}^0}$ for a given user $\vec{u}^0$ is $O(kn_m)$ from eq.~( \ref{eq:sugproc}) and \cite{fm1}. The complexity of sorting $N_m$ items is classically known as $O(N_m \log N_m)$. Hence to sort $N_m$ ratings for the user $\vec{u}^0$ with a trained FM, with size-$n_m$ vectors representing these items, requires the complexity $O(N_m\times kn_m + N_m \log N_m )$. 
%Since $k=O(1)$, the overall complexity of the suggestion process for a fixed user in the direct method is $O(n_mN_m^2\log N_m)$. 

Note that if the vectors $\vec{m}$ for encoding $N_m$ items are in binary form and denote $\log\equiv\log_2$, then $n_m \approx \log N_m$.  In this case, the complexity of the suggestion process becomes 
$$O(N_m\times kn_m + N_m \log N_m )\sim O(kN_m \log N_m),$$ 
where $k=O(1)$.
%%%%%%%%%%%%%%%%%%%

%%%%%%%%%%%%%%%%%%%%%%%%%%%%%%%%%%%%%%%%%%%%%%%%%%%%%%%
\subsection{Quadratic Unconstrained Binary Optimization}

\label{sec:qubo}
Quadratic unconstrained binary optimization (QUBO) is an NP-hard combinatorial problem that can apply to the traveler salesman problem, finance portfolio optimization, max-cut problem, and even machine learning \cite{toising2}. With binary variables $x_i \in \{0,1\}$, linear weights $w_i \in \mathbb{R}$, and coefficients $W_{ij} \in \mathbb{R}$, the equation of a QUBO problem is: 
\begin{eqnarray}
\label{eq:qubo}
y_{\text{QUBO}}(\vec{x}) := \sum_{i = 1}^n w_i x_i + \sum_{i < j}^n W_{ij} x_i x_j.
\end{eqnarray}
%\revise{negative value?}
The solution to a QUBO problem is a binary vector $\vec{x}^*$ that minimizes the objective value $y_{\text{QUBO}}$:
\begin{eqnarray}
\vec{x}^* = \arg \min_{\vec{x}} y_{\text{QUBO}}(\vec{x}).
\end{eqnarray}
Note that one can transform a minimization problem into a maximization problem by adding a minus sign to the objective value. Moreover, it is possible to map a  QUBO problem into an Ising problem \cite{toising1} with the relation $\sigma^z_i = 2x_i -1 $. In particular, a QUBO problem is computationally equivalent to the following form :
\begin{eqnarray}
\label{eq:ising}
H_p = \sum_{i=1}^n h_i \sigma^z_i + \sum_{i < j}^n J_{ij} \sigma^z_i \sigma^z_j,
\end{eqnarray}
where $h_i \in \mathbb{R}$ and $J_{ij} \in \mathbb{R}$ are the corresponding external field and coupling terms, and $\sigma^z$ is the Pauli-$z$ operator. Thus, we can solve a QUBO problem by solving its corresponding Ising form. This approach is achievable by energy minimization through adiabatic quantum computation, such as quantum annealing. 

\subsection{Quantum Annealing}
\label{sec:qa}

Quantum Annealing (QA), formulated in its current form by T. Kadowaki and H. Nishimori \cite{qa1}, is a meta-heuristic approach capable of solving combinatorial optimization problems by utilizing the quantum mechanical nature of the associated physical systems. The QA process evolves from the ground state $|\psi_i\rangle$ of an initial Hamiltonian $H_i$ to the ground state $|\psi_p\rangle$ of the problem Hamiltonian $H_p$ (eq.~(\ref{eq:ising})), where presumably one can easily prepare $(H_i, |\psi_i\rangle)$. Mathematically, this adiabatic evolution can be described by the Hamiltonian in the following form, with $s(t)$ evolving from 1 to 0 during a QA process:
\begin{eqnarray}
H(t) = s(t) H_i + (1-s(t))H_p.
\end{eqnarray}
For example, it is set to be $H_i \propto \sum_{i = 1}^n \sigma_i^x$ in D-Wave's implementation \cite{dwave_qa}. With a proper annealing schedule $s(t)$, probability of finding the ground state of $H_p$ is close to 1 \cite{qa2}. Thus, QA can search for the solution to a QUBO problem by studying the corresponding Ising form (eq.~(\ref{eq:ising})). This approach has a wide range of applications for different problems mentioned in Sec.~\ref{sec:qubo} \cite{toising2}.
%%%%%%%%%%%%%%%%%%%%%%%%%%%%%%%%%%%%%%%%%%%%%%%%%%%%%%%
\begin{figure}
\centering
    \includegraphics[scale=0.25]{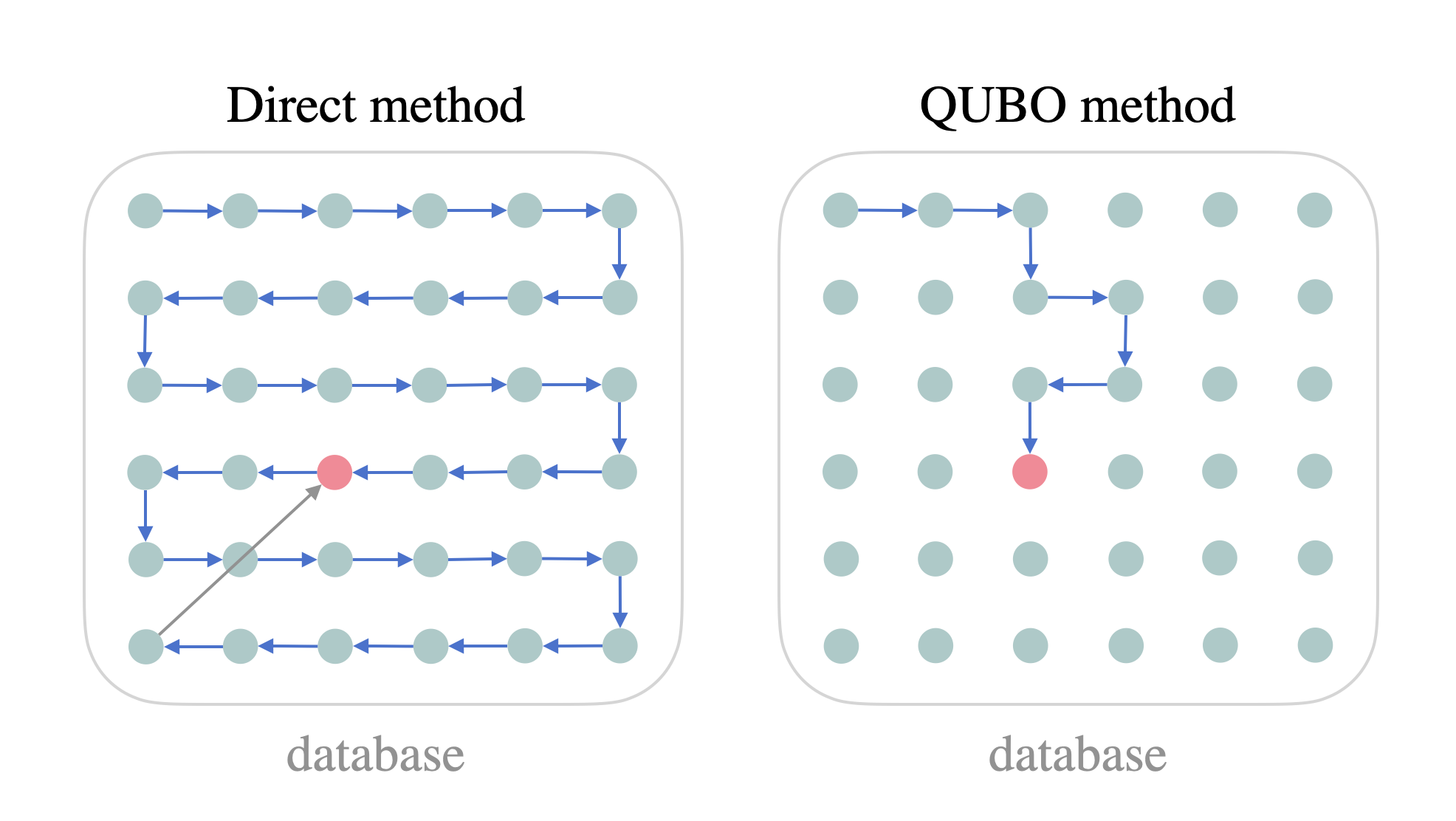}
    \caption{
    Schematic diagram. The direct method needs to predict the rating for every data point in the database and return the highest rating candidate at the end (red dot). In contrast, the QUBO method considers the database as the energy landscape of an energy minimization problem, which only calculates the data points along the optimization trajectory. A specific optimization method can significantly reduce the computational complexity (quantum annealing in our case).} 
\label{fig:schematic_overlap}
\end{figure}

%%%%%%%%%%%%%%%%%%%%%%%%%%%%%%%%%%%%%%%%%%%%%%%%%%%%%%%

%\section{Theoretical Result}
%\label{sec:theoretical result}
%In this section, we derive the QUBO formulation of the suggestion process for a fixed user from a trained FM and then estimate the computational complexity of such process solved by quantum annealing. 

\section{Trained Factorization Machine as QUBO}
\label{sec:tfmaq}
%\mh{summarizing paragraph.}
%\revise{problem description here, recommendation system, parameter setting}

In this section, we derive the QUBO formulation of the suggestion process with a given trained FM and a fixed user. The key component here is that we will specialize the $d$-dimension input vector $\vec{x} = (\vec{u}, \vec{m})$ in eq.~(\ref{eq:fm}) to be in the \emph{binary form} where $\vec{u}\in\{0,1\}^{n_u}$ encodes a user and $\vec{m}\in\{0,1\}^{n_m}$ an item in the dataset, with the dimension relation $n_u + n_m = d$.

Consider an FM recommendation system modeled by eq.~(\ref{eq:fm}), where the training process described in Sec.~\ref{sec:fm} determines the global bias $w_0 \in \mathbb{R}$, the weight $\vec{w} \in \mathbb{R}^d$ and the feature embedding ${\bf V} \in \mathbb{R}^{k \times d}$. When assuming the input vector $\vec{x}$ to be binary, we view eq.~(\ref{eq:fm}) as a QUBO problem equation eq.~(\ref{eq:qubo}) with $\vec{w}$ remaining the same and $W_{ij}=v_i^T v_j$. Note that eliminating the bias term $w_0$ does not affect the underlying optimization problem. Given the meaning of $\vec{x}$ in FM to be a pair of user and item data, and $y_{\text{fm}}$ the corresponding rating (or score), by introducing a negative sign to the associated QUBO problem:
\begin{equation}
    \label{fmqubo}
    - y_{\text{fm}}(\vec{x}) = y_{\text{QUBO}}(\vec{x}) = \sum_{i = 1}^d w_i x_i + \sum_{i<j}^d W_{ij} x_i x_j,
\end{equation}
the solution $\vec{x}$ now becomes the highest rating pair for the resulting minimization problem. 

To formulate the suggestion process for a picked user represented by the binary vector $\vec{u}^0$, we fix the user part $\vec{u}=\vec{u}^0$ in the input vector $\vec{x}=(\vec{u},\vec{m})$, which  reduces the dimension of the corresponding QUBO from $d$ to $n_m$:
% Divide the Ising model part to the next equation
% describe the process seperatetly
\begin{eqnarray}
&& y_{\text{QUBO}}(\vec{x})|_{\vec{u} = \vec{u}^0} = \sum_{i = 1}^d w_i x_i + \sum_{i<j}^d W_{ij} x_i x_j \nonumber \\
 = &&\ \sum_{i = 1}^{n_u} w_i u_i^0 + \sum_{i = 1}^{n_m} w_{n_u + i} m_i \nonumber \\
&& + \sum_{i < j}^{n_u} W_{ij} u_i^0 u_j^0 + \cdots + \sum_{i<j}^{n_m} W_{i+n_u, j+n_u} m_i m_j.
\end{eqnarray}
By summing over like terms, we further reduce this equation to
\begin{eqnarray}
\label{eq:sugproc}
%&& y_{\text{QUBO}}(\vec{x})|_{\vec{u} = \vec{u}^0} \nonumber \\
&& = \sum_{i = 1}^{n_m} \tilde{w_i} m_i + \sum_{i<j}^{n_m} \tilde{W}_{ij} m_i m_j + \text{offset} \nonumber \\
&& \equiv \sum_{i = 1}^{n_m} \tilde{w_i} m_i + \sum_{i<j}^{n_m} \tilde{W}_{ij} m_i m_j, 
\end{eqnarray}
where the ``offset'' term in the last step is neglected as it does not affect the underlying optimization problem. Finally, we can map eq. (\ref{eq:sugproc}) into an Ising model on the $n_m$-dimensional lattice as discussed in Sec. \ref{sec:qubo}:
\begin{eqnarray}
\label{eq:final}
&& y_{\text{QUBO}}(\vec{x})|_{\vec{u} = \vec{u}^0} = \sum_{i = 1}^{n_m} \tilde{h_i} \sigma_i + \sum_{i < j}^{n_m} \tilde{J}_{ij} \sigma_i \sigma_j.
\end{eqnarray}
By calculating the ground states and possibly some low-lying excited states of the Ising problem with QA or other algorithms, we can transform the quest of producing high-rating recommendations for the trained FM into the task of finding low-lying energy states for the corresponding Ising problem. 

\section{Experimental Results and discussion}
%paragraph 
% 1. setup 
% 2. parameters 
% 3. result
\label{sec:experimental result}
We use the well-known MovieLens-20M dataset \cite{movielens20m} to demonstrate the effectiveness and behaviors of our hybrid algorithm, with $\vec{u}$ and $\vec{m}$ the binary encoding of userID and movieID, and $y_{\text{fm}}(\vec{x})$ the prediction of rating. Note that even though FM can utilize side information of users and movies, here we only use userID and movieID to train an FM. The dataset  MovieLens-20M contains 20 million ratings of 27000 movies by 138000 users, and we use different fractions of this dataset and up to 6 million ratings to observe behaviors of our method.

% Movielens20M : 20 million ratings and 465,000 tag applications applied to 27,000 movies by 138,000 users. 

\begin{figure*}
\centering
    \includegraphics[scale=0.3]{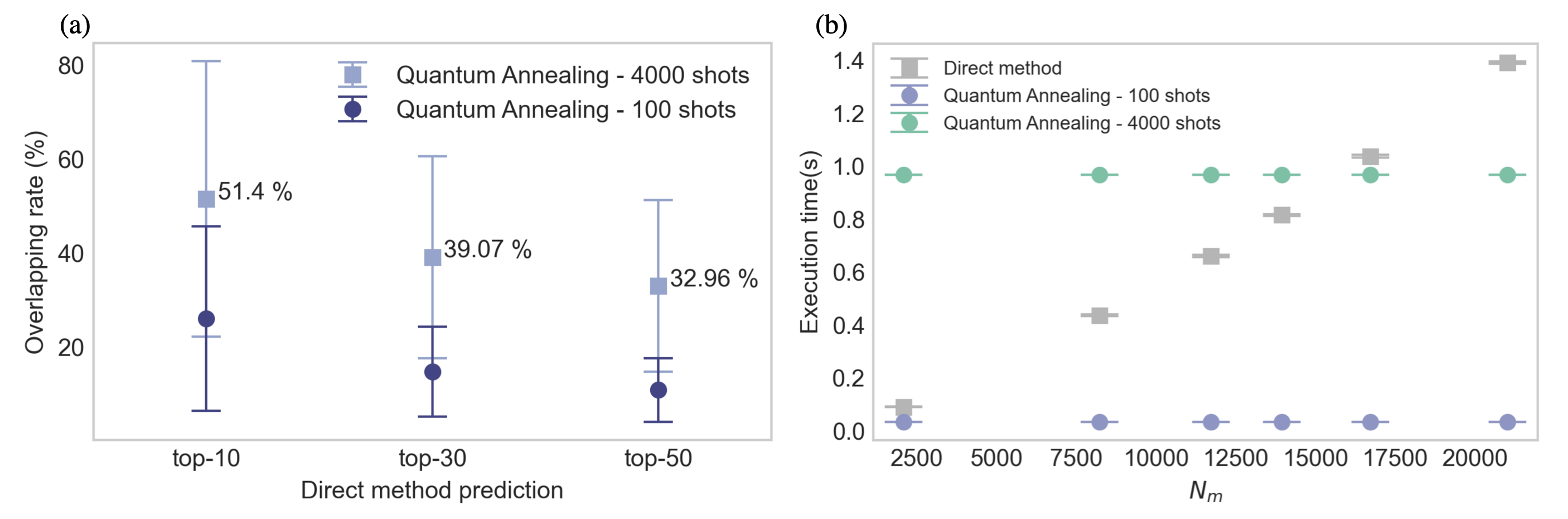}
    \caption{(a) Overlapping ratings between predictions from direct method and QA for both 100 measurement shots and 4000 measurement shots. The FM is trained by 6 million ratings for 41305 users and 21011 movies ($N_{\text{data}} = 6 \times 10^6, N_u = 41305, N_m = 21011, n_u = 16, n_m = 15$). (b) Execution time for both direct method and QA with $N_{\text{data}} \in \{5 \times 10^3, 10^5, 4 \times 10^5, 10^6, 2 \times 10^6, 6 \times 10^6 \}$, $N_m \in \{2090, 8227, 11719, 13950, 16715, 21011\}$ and $n_m \in \{12, 14, 15\}.$
    } 
\label{fig:qa_result_1}
\end{figure*}

\subsection{Set Up}
In our experiment, we first train an FM model with latent factors of dimension $k=200$ and then transform it into the corresponding Ising problem for the suggestion process. 

However, there is an issue with binary encoding: the dimension of the solution space in quantum computing is typically more than the number of items (all possible movies for us) from the dataset. For the direct method, the solution space of the suggestion process is the number $N_m$ of all the movies, while the dimension of the solution space for the QUBO method is $2^{\lceil \log_2 N_m \rceil}$. In this case, QUBO method may consider $2^{\lceil \log_2 N_m \rceil} - N_m$ binary vectors that is not in the movie list as solutions. 
To deal with this issue, we map all the computational basis vectors of the solution Hilbert space to our movie list (item dataset) by encoding the $2^{\lceil \log_2 N_m \rceil} - N_m$  higher-rating movies (items) in the dataset to $2$ different binary vectors. Thus, any solution sampled by quantum annealing will produce one of the existing movies.

The FM in this case is trained by 6 million ratings for 41305 users and 21011 movies, so that the lengths for the user and movie part of the input vector are respectively  $n_u = \lceil \log_2 41305 \rceil = 16$ and $n_m = \lceil \log_2 21011 \rceil = 15$. In particular, the QUBO/Ising problem size of the suggestion process for a fixed user is 15. We use the quantum annealer device D-Wave \textsf{DW\_2000Q\_6}, which has 2000 qubits and 6000 couplers. 

\subsection{Execution time of Quantum Annealing}
 By using different fractions of MovieLens-20M dataset as training data, Fig.~\ref{fig:qa_result_1}(b) shows the execution time for direct method and QA for $N_m \in \{2090, 8227, 11719, 13950, 16715\}$, where each data point represents the average results of 5 randomly chosen users. We perform QA on the D-Wave \textsf{DW\_2000Q\_6} QPU and the direct method on an Apple \textsf{M1 Max} chip with a 10-core CPU. Note that the upper limit of total execution time for quantum annealing is set to 1 second in the platform we used.

The execution time shows that QA is an efficient sampling method for suggesting candidates in the FM recommendation system. With some sacrifice of solution quality (Fig.~\ref{fig:qa_result_1}(a)), this could be useful if $N_m$ is extremely large and there is no efficient algorithm in the classical approaches.

\subsection{Solution Quality}

We use the \emph{overlapping rate} (in percentage) as the standard of our search  results. Here the overlapping rate in the top-$k_s$ experiment is defined as the proportion of QA-captured overlapped movies in the top-$k_s$ list from the direct method.  Fig.~\ref{fig:qa_result_1}(a) shows the overlapping rate of QA movie recommendation for both 100 measurement shots and 4000 measurement shots. 

In Fig.~\ref{fig:qa_result_1}(a), each data point  represents the average of QA suggesting results from 100 randomly picked users, while a  vertical strip represents the overlapping rate of each QA 4000/100-shot results for $k_s \in \{10, 30, 50\}$. We observe that QA captures only a small amount of targeting movies from the 100-shot result but relatively more movies from the 4000-shot result. 

On average, the overlapping rates are $51.4 \%$, $39.07 \%$, and $32.96 \%$ from the top-10, top-30, and top-50 suggestions from the 4000-shot results. The overlapping rate is higher for a smaller $k_s$, which is consistent with the principle that QA samples are more frequently lowly-excited states.  

To the best of the authors' knowledge, we can not use a typical Adiabatic Theorem (AT) to estimate our overlapping rates date. Denote $H_{t_f}(t)$ the Hamiltonian governing a quantum annealing process with total evolution time $t_f$ and $H(s)=H_{t_f}(st_f)$ the normalized Hamiltonian with the dimensionless parameter $s=t/t_f\in[0,1]$. A typical AT asserts roughly that for all $s\in[0,1]$
\begin{eqnarray}
\label{eq:AT} \|P_{t_f}(s)-P(s)\|\leq O(\frac{C_H}{t_f\Delta^3})
\end{eqnarray}
where $P_{t_f}(s)$ and $P(s)$ are the spectral projections at time $s$ to the eigenspace of the lowest $m$ eigenvalues, with the spectrum separated by a gap \mbox{$\Delta=\min_s\{\Delta(s)\}>0$}, and $C_H$ is a constant involving smoothness of $H(s)$,
cf. \cite[Section II]{qa2}. Consequently, one can conclude that the result of QA approximates nicely to the lowest $m$ eigenstates when $t_f\gg0$. But note that $t_f=1$ second in the reality of applying the D-Wave platform, where the upper bound in eq. (\ref{eq:AT}) is always larger than one and provides no information.

Due to the limitation of the execution time (=1 sec.), this brings up the necessity again to improve eq. (\ref{eq:AT}). Our experiment provides possible evidence that the upper bound in eq. (\ref{eq:AT}) under suitable conditions, as in the applications to the recommendation systems, can be significantly improved.

\subsection{NISQ Advantage of Quantum Annealing in FM recommendation system}

In the realm of quantum annealing, the ultimate goal is to achieve the ground state of the Hamiltonian. This state represents the optimal solution sought after for the given problem. However, noise and the inherent constraint of annealing time on present-day NISQ hardware pose significant challenges. Consequently, quantum annealing often generates a diverse set of sub-optimal solutions instead of obtaining a single, flawless solution.

Interestingly, these sub-optimal solutions align with the fundamental objectives of recommendation systems. Rather than merely providing the absolute best solution, recommendation systems are designed to present multiple suggestions to users. Although solutions generated by quantum annealing in NISQ hardware are not the absolute best, the sub-optimal ones offer a valuable range of choices for the recommendation system to propose to users. This assortment of suggestions can cater to individual preferences, diverse contexts, and the inherent subjectivity often present in decision-making processes.

Therefore, despite the inherent limitations and imperfections introduced by noise and annealing time constraints in NISQ hardware, the sub-optimal solutions provided by quantum annealing are unexpectedly well-suited for fulfilling the objectives of recommendation systems. This convergence of goals opens up exciting possibilities for leveraging quantum annealing in recommendation systems.

% D-Wave

% \begin{figure}
% \centering
%     \includegraphics[scale=0.19]{scaling.png}
% \caption{Scaling behaviors of the execution time for both direct method and QA in the case of $N_m$ from $10^1$ to $10^{47}$. Corresponding QUBO sizes $(\lceil \log_2 N_m \rceil)$ of different $N_m$'s are also plotted, with indications of the problem size upper limits for D-Wave's current hardware.
%     } 
% \label{fig:scaling}
% \end{figure}

%$N_{\text{data}} \in \{5 \times 10^3, 10^5, 4 \times 10^5, 10^6, 2 \times 10^6, 6 \times 10^6 \}$

\begin{figure*}
\centering
    \includegraphics[scale=0.30]{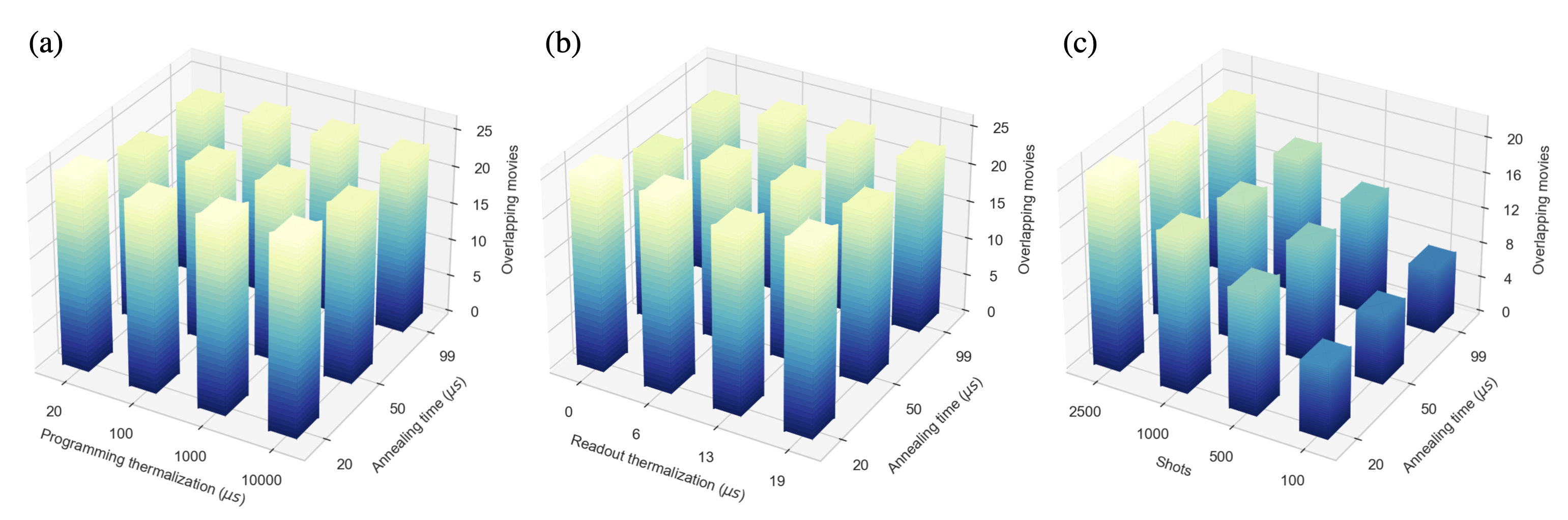}
    \caption{The number of overlapping movies between top-100 suggestions from the direct method and QA for different annealing parameters. (a) Different programming thermalization and annealing time with measurement shots $=2500$ and readout thermalization $= 0~\mu \text{s}$. (b) Different readout thermalization and annealing time with measurement shots $=2500$ and programming thermalization $= 1000 ~\mu \text{s}$. (c) Different measurement shots and annealing time with readout thermalization $ = 0~\mu \text{s}$ and programming thermalization $= 1000~\mu \text{s}$.  
    } 
\label{fig:annealing_para}
\end{figure*}

% \subsection{Scalability}
% Our experiment includes examples of dataset size $N_{\text{data}} \in \{5 \times 10^3, 10^5, 4 \times 10^5, 10^6, 2 \times 10^6, 6 \times 10^6 \}$ with the corresponding QUBO size $n_m \in \{12,14,15 \}$.  

% In Fig,~\ref{fig:scaling}, we extrapolate the fitting curves in  Fig.~\ref{fig:qa_result_1} up to  $N_m \sim 10^{49}\sim 2^{163}$ to see the execution time scaling in the extreme case. To visualize the capability of the current D-Wave quantum annealers, we also include the estimation of the corresponding QUBO size upper limits in the graph. For the 5760-qubit D-Wave Advantage 4.1 system, the upper limit for a general Ising problem is around 145 due to the non-fully connectedness of current hardware. For the same reason, this upper limit for the 2000-qubit D-Wave DW\_2000Q system is 64 \cite{dwave_upper_limit}. 

% We can observe that the fitting curve related to QA is dozens of orders less than that of the direct method near the case of the D-Wave upper limit. The ``QUBO size" axis of the figure can also be considered as the length of the encoded binary vector of the items in the problem formulation and is not necessarily the same as our MovieLens example. In particular, it can include more information besides item identities, as in our case. 

\subsection{Effects of Annealing Parameters}
Since QA involves several factors that could affect our experiment, to understand our algorithm's properties better, we investigate different tunable parameters of the QA process to see the behavior of the results, where the energy scale of the quantum annealing process is in the order of $10^{-24}$ joule according to D-Wave ~\cite{dwave_energy_scale}. We select four tunable parameters to investigate:  ``Annealing time'', ``Programming thermalization'', ``Readout thermalization'' and ``Shots''. According to the solver parameters document from D-Wave ~\cite{dwave_solver_para}: ``Annealing time'' is the time duration of the quantum annealing; ``Programming thermalization'' is the time to wait after programming the quantum annealer for it to cool back to a base temperature; ``Readout thermalization'' is the time to wait after each state is read from the quantum annealer for it to cool back to the base temperature;  ``Shots'' is the number of states to read from the solver. We use the number of overlapping movies between top-100 suggestions from the direct method and QA to analyze the performance.

Fig.~\ref{fig:annealing_para}(a) investigates different parameters of programming thermalization and annealing time, where the number of overlapping movies shows no significant difference among these variations of parameters. One can also observe similar behavior in Fig.~\ref{fig:annealing_para}(b): the readout thermalization and annealing time could result in the thermal noise in the system, which is in the same order of magnitude as the thermal variation before and after the waiting time. For annealing time, these results show that the default value 20 $\mu \text{s}$ is already sufficient for our QA process. Thus no improvement would be found with larger values. In Fig.~\ref{fig:annealing_para}(c), it is no surprise that results with more shots perform better since we have more chances to hit the high score candidates in these cases.  

%%%%%%%%%%%

\section{CONCLUSION AND FUTURE WORK}
\label{sec:conclusion}
 
In this work, by transforming a trained FM from user dataset into a QUBO formulation, we construct a hybrid recommendation system based on solving the associated Ising problem via quantum annealing.

Theoretically, we first derive the QUBO formulation for the suggestion process and then describe the potential advantage of NISQ hardware quantum annealing in recommendation systems.
% prove that the computational complexity has an faster than quadratic speedup improvement: $$N_m^2 \log N_m \rightarrow e^{\sqrt{\log N_m}}$$ compared to the classic direct method.

Experimentally, we apply our hybrid algorithm to the MovieLens-20M dataset to demonstrate the applicability of current NISQ hardware and analyze the solution quality. By encoding the whole solution space of QUBO to the movie list, we use the D-Wave quantum annealer to solve the corresponding optimization problem of the suggestion process. Our experiment shows that the top-10 suggestion has $51.4 \%$ average overlapping rate with that from the direct method and has the scaling behavior of the execution time matching our theoretical results. 

Our future work could be searching for other use cases of FM and studying whether the quantum annealing approach is advantageous. Moreover, the limitation of the execution time of the current NISQ hardware leads to the necessity of further developing a finer math theory for improving the Adiabatic Theorem, which under suitable conditions, applies to the recommendation systems. Our numerical experiment provides evidence for the existence of such a theorem.

% reason of writing this ? 
\section{Acknowledgments} Hsin-Yu Wang, Pei-Yen Liao and Ching-Jui Lai acknowledge the funding support from the Ministry of Education in Taiwan for developing an innovative training scheme to bring intelligent  students to frontier research. We thank Ming-Da Liu and Chiao-Ching Huang from the Data Management department of Cathay Life Insurance in Taiwan for bringing up this question, and Simon Lin for valuable discussions. Finally, we thank the Quantum Technology Cloud Computing Center, National Cheng Kung University, Taiwan, for supporting our  access to D-Wave's QPU through Amazon Braket.

% The \nocite command causes all entries in a bibliography to be printed out
% whether or not they are actually referenced in the text. This is appropriate
% for the sample file to show the different styles of references, but authors
% most likely will not want to use it.
\nocite{*}

\bibliography{references}% Produces the bibliography via BibTeX.

\end{document}